\documentclass[preprint,showpacs]{revtex4}

\usepackage{graphicx}
\usepackage{amsmath}
\usepackage{bm}

\begin{document}

\title{Low frequency excitations in the Coulomb Glass: numerical analysis using the avalanche method}

\author{A.V. Shumilin}  \email{AVShumilin@mail.ioffe.ru}
\affiliation{A.F. Ioffe Physico-Technical Institute, St.-Petersburg
194021, Russia}

\date{\today}

\begin{abstract}
We develop a numerical method, which  allows to find pairs of metastable states differing by transition of several electrons. We
show that at low temperature these pairs can be treated as local metastable systems that determine low-frequency properties of Coulomb Glass
with low external disorder. The contribution of these pairs to low-frequency properties is suppressed when the strength of the external disorder becomes comparable with interaction between neighboring electrons.
\end{abstract}


\maketitle

\section{Introduction}

The conductivity of disordered system with localized electrons is governed by phonon-assisted hops between localized states. It is known that at low temperature the characteristic lengths of the hop increase with decreasing temperature leading to a phenomenon known as variable-range hopping. It was first understood by Mott \cite{Mott}, who derived the law named after him for the temperature dependence of the conductance:
\begin{equation}\label{Mott}
\sigma \propto \exp \left[ - \left( \frac{T_0}{T} \right)^{1/(d+1)} \right],
\end{equation}
where $d$ is the system dimensionality, $T$ is temperature and $T_0$ is a material-dependent constant. The major assumption that leads to the Mott law is that the density of states at the Fermi level is finite at zero temperature.

Later it was understood that the physics of variable-range hopping is strongly influenced by long-range Coulomb interaction between localized electrons. The most known result of this interaction is that the density of states at the Fermi level is equal to zero at $T=0$ due to formation of so-called Coulomb Gap. It leads to another temperature dependence of conductivity at low temperatures (for details see \cite{Efr-Sh})
\begin{equation}\label{Mott}
\sigma \propto \exp \left[ - \left( \frac{T_1}{T} \right)^{1/2} \right].
\end{equation}

It is believed however that effects of Coulomb interaction are not limited to formation of the Coulomb Gap. The interplay of long range interaction and disorder leads to the glass-like behavior of a system with localized electrons. For this behavior the discussed systems are sometimes called electron or Coulomb Glasses. Coulomb Glasses show a variety of properties
related to long relaxation times (or, that is the same, low frequencies) that are similar to normal (structural) glasses. Among these properties are $1/f$  noise \cite{exp1f,exp1f-2} and slow relaxation observed in a number of experimental works \cite{expOva, Ova-dyn, Ova-dis, Ova-Ti2O3}.

However up to now there is no well-accepted theory of these properties.  Shklovskii \cite{Shklovskii-pores} showed that $1/f$ noise can be described in terms of one-electron physics, but only in limited frequency range determined by the same constant $T_0$ that governs temperature dependence of conductivity. In numerical studies \cite{Amir1, Amir2} the authors found that slow relaxations can also be described with one-electron hops in a limited frequency range. Although the frequency limits were not discussed in \cite{Amir1, Amir2} we believe them to be the same as in \cite{Shklovskii-pores}. Thus, although the importance of many electron hops to dc conductivity is still under discussion \cite{2e-1,2e-2,2e-3,2e-4}, it is widely believed that low-frequency properties out of this range should be ascribed to many-electron physics.

 In \cite{Kogan} it was shown that due to electron-electron (Coulomb) interaction the ground state of a system containing interacting localized electrons can hardly be reached. The system will rather freeze into one of metastable states (so-called valleys). A transition between these valleys at low temperature can occur only through simultaneous tunneling of a large number of electrons and thus a transition time can be indefinitely large. The main idea of \cite{Kogan} was to simulate the Coulomb Glass considering a numerical sample with localized electron states with some extent of disorder. The authors started with random distribution of electrons over localized states (sites) and then resolved all one-electron hops that decrease the total energy of the system. All the states of the sample (i.e., distributions of electrons over sites) that could not provide one-electron hops decreasing the total energy were considered as valleys.

The transitions between valleys can go either with thermal activation or through the simultaneous hops (co-tunneling) of many electrons. The probability of activation exponentially decreases with decreasing temperature while the probability of co-tunneling (although initially small) does not depend on temperature. So one may conclude that at sufficiently low temperatures the transitions between valleys are possible only due to many-electron hops.

Later algorithms, analogous to \cite{Kogan}, were developed \cite{Gal-Bergly-aging,correlation,Somoza,Katzgraber, Goethe0} allowing consideration of electron glass at finite temperature and description of its dynamics due to one-electron hops.  However, all such algorithms do not include many-electron hops and thus cannot describe transitions between the valleys at low temperature.

There were other attempts to describe slow relaxations of Coulomb Glass. In \cite{Efr-Ts} authors considered the difference between conductivities of Coulomb Glass in different metastable states. They showed that it is comparable with the experimentally observed variation of conductivity due to slow relaxation when they included no external disorder in the simulated Coulomb Glass. With external disorder, the system still possessed metastable states but the difference between conductivities of these states was found to be negligible. However the characteristic times of transitions between valleys and thus characteristic times of the relaxation were out of the scope of that work.

Another attempts to describe low-frequency properties of Coulomb Glass were made in \cite{agr1f,agrrel, agrOur0, agrour}. The authors assumed that the Coulomb Glass contains local areas that can exist in one of two states with close energies and act as local metastable systems. It was assumed that a transition between these states involves several electrons (depending on the metastable system size) and can occur either by co-tunneling of all the electrons or by thermal activation and the formation of a domain wall between regions with different coordinations of occupied and empty sites.  In both cases the relaxation time at low temperatures exponentially grows with system size. These systems can act as slow traps for electrons near the Fermi level and thus modify the conductivity.

It turns out that traditional numeric experiments (like \cite{Kogan}) cannot detect such systems. One of the reasons for this may be as follows. In the discussed computations one places electrons randomly on the localized sites and then simulates hopping until some state that can be called metastable is reached. In \cite{Kogan}  it is the state that is stable with respect to all transitions of one electron (1e stable state). Then electrons are placed in new random positions and the calculations repeated. In the same manner one obtains a set of metastable states. However, there is no guarantee that all metastable states are found. Therefore one cannot be sure whether 1e stable states that differ only locally exist.

An alternative method is based on the response of a Coulomb Glass to an external perturbations. In \cite{Shk-screen} the numerical study of a response of an insulator with hopping conductivity to external electric field was provided to understand the nature of screening in these systems. In \cite{Goethe} the authors studied the response to a local excitations. They inserted an additional electron into the glass (charge excitation) or moved an electron to another place (dipole excitation) and then considered its impact on the Coulomb Glass. It occurred that rather often such an excitation led to so-called avalanches, i.e., after the excitation the state of the rest (unchanged) part of the Coulomb Glass appeared to be 1e unstable and there existed one-electron hops that decreased total energy. The avalanches themselves are the series of these hops leading the Coulomb Glass (with the exception of the initial excitation) to a new 1e stable state. Some avalanches were found to be quite large (comparable to the size of simulated sample). Very recentness analogous algorithm was applied to study avalanches induced by external electric field \cite{K-arx}.

In our opinion, these avalanches have some similarities with metastable systems discussed in \cite{agr1f,agrrel, agrOur0, agrour}. However, the final state of the avalanche in \cite{Goethe} is not always a real 1e stable state of the whole Coulomb Glass (at least for dipole excitations) due to initial excitation that is still present in the system. After the dipole excitation authors resolved all 1e hops other than the hop that would reverse the initial excitation. Therefore, it is not possible to  say if after the avalanche the hop that reverses the initial excitation will increase or decrease the total energy and if the state of numerical sample after the avalanche in \cite{Goethe} is truly the 1e-stable state.

The goal of the present study is to modify the avalanche method so that the final state of each avalanche can be considered as a metastable state of the Coulomb Glass and then use this method to analyze low-frequency properties of the Coulomb Glass.

\section{Avalanche calculation}

To simulate the Coulomb Glass we consider a square numerical sample containing $N_S$ localized electron states (sites) with $N_S/2$ electrons. The sites are distributed randomly over the sample. The length units  were selected in such a way that the concentration of sites $N_S/L^2 = 1$ ($L$ is the side of the numerical sample). The energy of the system in a given configuration (i.e., with a given filling numbers) is as follows \cite{Efr-Sh}
\begin{equation}
E = \sum_i U_i n_i + \sum_{i \ne j} \frac{e^2(n_i-q_0)(n_j -q_0)}{r_{ij}}, \quad -U<U_i<U.
\end{equation}
Here $n_i$ is the filling number of site $i$, $U_i$ is the random energy associated with the site. It is assumed to be uniformly distributed between values $-U$ and $U$ (in the first part of our work $U=0$), $r_{ij}$ is the distance between sites $i$ and $j$, and $q_0$ is the background charge introduced to keep electrical neutrality. We consider numerical samples with $N_S = 5000$ and $2500$ electrons. The electron charge $e$ was considered to be unity (that means that the energy unit is the Coulomb interaction of two electrons at the distance $r=1$). We use cyclic boundary conditions with cutoff at distances larger than $L/2$.

The first step of our simulation is to obtain the state of the sample that is the local energy minimum with respect to all one-electron hops (1e hops). To find this state we start with random distribution of the electrons over the sites and then resolve all 1e hops that lower the total energy of the system with the optimized Monte-Carlo algorithm described in details in \cite{Galperin-Algorithm}. The important part of this algorithm is division of all hops into the hops between nearest neighbors and the hops between more distant sites. The work \cite{Galperin-Algorithm} considers Coulomb Glass on a lattice, so the division is straitforward. Here we deal with Poissonian distribution of impurities (that allow us to consider Coulomb Glass with low energy disorder), so we introduce the artificial network of neighbors. The states are considered as neighbors when they are closer than $r_{nn} = 2L/\sqrt{\pi N_S}$. The choice of $r_{nn}$ is arbitrary, this parameter impacts only on the computation time and not the result of the simulation --- all the hops both between neighbors and not between neighbors are resolved at the end of this step. The result of this algorithm is the state of the Coulomb Glass that is stable to all hops of a single electron.

The second important step is calculation of the avalanches induced by dipole excitations. Recently it was shown \cite{Goethe} that introducing a dipole excitation into a Coulomb Glass (e.g., moving an electron to a neighbor free site) can lead to an avalanche --- the appearance of new possibilities to hop with a decrease of total energy. The realization of these possibilities can lead to new energy-decreasing hops and so on. Thus, a local excitation in a Coulomb Glass can lead to a change of the state of some part of the glass. The distribution of these avalanches was found in \cite{Goethe} to be quite wide: from situations when there is no avalanche at all to the avalanches with sizes comparable with whole numerical sample.

The avalanches can be compared with local two-level systems (chessboard clusters) considered in \cite{agr1f,agrrel, agrOur0, agrour}. These cluster are in some way analogous to a part of chess board, each cell corresponds to a localized electron state. The cluster has two metastable states. In one state all ``black" sites contain an electron while all ``white" sites are empty. In the second state ``black" sites are empty while ``white" sites are filled. When in such a cluster (in its 1-st metastable state) one moves an electron to a neighboring site, this electron appears in the place corresponding to the second metastable state (i.e., on  a``white" site). If the electron is not allowed to return, the rest part of the cluster will also eventually relax to the second metastable state. So if the analogs of such chessboard clusters exist in a Coulomb Glass, then the avalanche calculation can help to ``detect" them. Here we use the modified avalanche calculation to detect the places that can be similar to these clusters and then try to answer two questions. The first question is: can these clusters be considered as independent from the rest part of the Coulomb glass, and the second one is: can they significantly modify the low frequency properties that can be obtained from 1e hops \cite{Shklovskii-pores}.

Let us note that (as it was discussed above) the state of the system that is obtained after avalanche calculation in \cite{Goethe} is not necessarily a metastable one. For this  reason the avalanche method \cite{Goethe} should be modified. Here we introduce the following modification of the avalanche method. We start with a metastable state of a Coulomb glass, then at the first step  we introduce a dipole excitation --- move an electron from site $i$ to a neighbor free site $j$. At the second step, we make all possible one-electron hops that do not include sites $i$ and $j$ involved in the initial dipole excitation. At the third step, we again make all possible one-electron hops, but now we include sites $i$ and $j$ in our computation. By definition, the final step of our algorithm yields the state of the Coulomb Glass that is stable for all 1e hops and can be considered as metastable.

It can occur that after such calculation the system returns to the same state it was before the avalanche. In this case we state that the given dipole excitation has not created an avalanche, or, in other words, the size of created avalanche is $0$. Quite often, however, the initial excitation leads to an avalanche with finite size $N$ (it means that filling numbers of $N$ sites have been changed due to the avalanche).

We use our avalanche algorithm in two parts of our work. Firstly, we use it to simulate additional relaxation of Coulomb Glass that is out of the scope of simple 1e hops. To do so we calculate avalanches induced by a randomly selected dipole excitation. Then we compare  the energies of the initial state and the final state after the avalanche. If the energy of the final state is lower we change the state of the Coulomb glass to the final state. In this way we resolve some number, $N_{Relax}$, of random dipole excitations (this number will be discussed later).

\begin{figure}[htbp]
    \centering
        \includegraphics[width=0.9\textwidth]{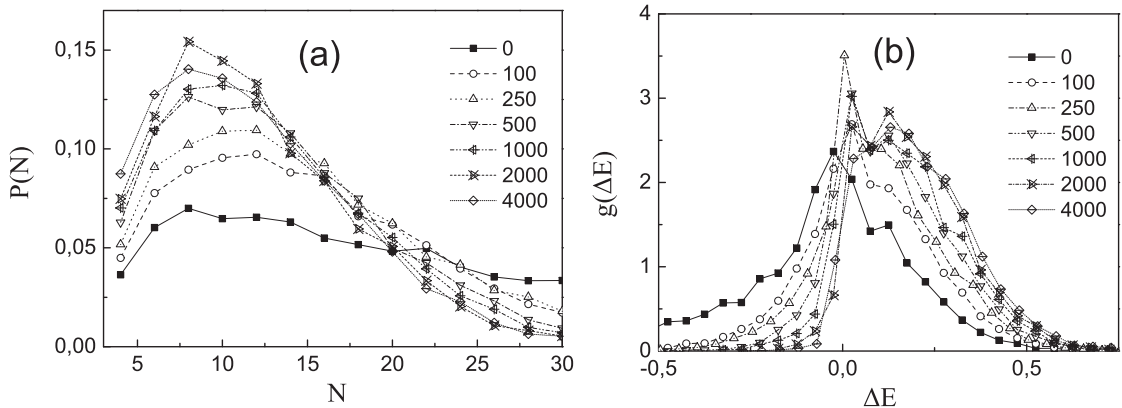}
        \caption{The distribution of avalanche sizes (a) and energies $\Delta E$ (b) for different numbers $N_{Relax}$. Numbers in the legend correspond to $N_{Relax}$.}
    \label{fig:AvSE}
\end{figure}

Then in the obtained state we calculate all avalanches induced by all possible dipole excitations (we, however, consider in this work only excitations that move an electron to a neighbor site). After each avalanche we return the system to its initial state. For computed avalanches we consider (at this point) two characteristics --- the size of the avalanche, $N$, and the difference in the total energy of the Coulomb Glass before and after the avalanche, $\Delta E$. Figure \ref{fig:AvSE} shows the distributions of these characteristics for different numbers  $N_{Relax}$. It can be seen that in the absence of initial relaxation the distribution of avalanches over energies $\Delta E$  is nearly symmetric. However the distribution for negative sign of $\Delta E$ dies out during the relaxation. Naturally, in the true ground state of the Coulomb Glass all avalanche energies $\Delta E$ should be positive. Also, the relaxation reduces the characteristic size of an avalanche. In the following computation we will always consider numerical samples with $N_{Relax}=2000$ (at this $N_{Relax}$ the distributions for $N$ and $\Delta E$ seem to be sufficiently relaxed).

The discussed computation shows that there are quite a lot of realizations of pairs of metastable states that differ only locally and thus are in some way similar to metastable clusters that are discussed in \cite{agr1f,agrrel, agrOur0, agrour}. To check this similarity we should answer two questions. (I) can these systems dominate the low frequency properties (at least in some frequency range). (II) can these systems be considered as independent from the rest part of the Coulomb glass at least at some temperatures.

\section{Avalanches and $1/f$ noise}

The second major step of our program is to associate our avalanches with local metastable systems considered in \cite{agr1f,agrrel, agrOur0, agrour} and understand what impact they can have on low frequency properties of the Coulomb glass. We will focus on low-frequency electrical noise in this work. However, it is known \cite{agrrel} that if the Coulomb glass has metastable systems responsible for $1/f$ noise, these systems can also lead to slow relaxation phenomena.

Let us consider the set of independent two-level systems, each one has the energy difference between its levels $\Delta E_i$, relaxation rate $\nu_i$ and changes the conductivity of the system by the value $\Delta \sigma_i$ (we consider these changes to be additive). These metastable systems result in the following conductivity noise \cite{Kogan-book}
\begin{equation}
\left( \sigma^2 \right)_\omega = \sum_i \left( \Delta \sigma_i \right)^2 w_T(i) \frac{\nu_i}{\nu_i^2 + \omega^2}, \quad w_T(i) = \frac{1}{ \cosh^2(E_i/2T)}.
\end{equation}
To understand the noise type one should find the distribution of logarithms of relaxation rates $\ln(\nu_i)$ with weights equal to $\left( \Delta \sigma_i \right)^2 w_T(i)$. The strong low-frequency noise will then correspond to a flat distribution of $\ln(\nu_i)$ in the corresponding frequency range.

In what follows we will associate computed avalanches with local metastable systems. We will try to answer which of the avalanches correspond to independent systems. For these systems we will find relaxation rates $\nu_i$ and ``temperature weights" $w_T(i)$. We will not be able, however, to estimate ``conductivity weights" $\Delta\sigma_i^2$. However, if the conductivity weight is not strongly correlated with relaxation rate one can find the distribution of $\ln (\nu_i)$ only with weights $w_T(i)$ (later on we call this distribution $W(\ln(\nu))$) and still understand the noise type: the flat distribution will indicate strong low-frequency noise while the white noise will correspond to rapid decrease of the distribution at low rates $\nu$.

 To associate an avalanche with a metastable system we do as follows. We consider all electrons that do not participate in the avalanche to be fixed in their places. However, for sites that took part in the avalanche we consider all possible states (arrangements of electrons over considered sites). The number of these states, $N_{st}$, grows very fast with the size of the avalanche: the number $N_{st}$ for an avalanche containing 2n sites and n electrons is $C_{2n}^n$. However, the relaxation time should grow exponentially with the size of the system, and thus even relatively small avalanches can yield relaxation times much larger than times of 1e hops. Here we are able to do the necessary calculation only for the avalanches that include no more than ten sites.

For every obtained metastable system (corresponding to some avalanche) we  calculate all the states $n$ of the system and their energies $E_n$. Then we calculate all the probabilities of direct transitions $t_{nm}$ (due to co-tunneling of all electrons participating in the transition) from any state $n$ to any other state $m$. For these probabilities we adopt the simplified model that takes into account only the exponential part of the dependence on the positions of sites. This model yields
\begin{equation}
t_{nm} = \exp\left( -\frac{R_{\sum}}{a} \right)\cdot \exp\left( - \frac{E_{ph}}{T} \right).
\end{equation}
Here $R_{\sum}$ is the minimal total distance that electrons in the system should cover when the system changes the state from $n$ to $m$. The minimum is taken over all possible combination of destination states for tunneling electrons. For example if the tunneling occurs from sites $1$ and $2$ to sites $3$ and $4$ then $R_{\sum}$ should be equal to the minimum
$R_{\sum} = \min(r_{13} + r_{24}, r_{14} + r_{23})$. $a$ is twice the localization radius, for our computation we take $a=0.33$. $E_{ph}$ is the energy of a phonon that should be absorbed during the transition. If $E_n < E_m$ then $E_{ph} = E_m - E_n$. If $E_n > E_m$ then the transition can go with phonon emission and we take $E_{ph} = 0$. $T$ is the temperature in energy units.

To calculate the relaxation rate let us consider the set of identical metastable systems. Let them have the probabilities $p_n$ to be in the state $n$. The relaxation then can be described by the set of differential equations
\begin{equation}
\frac{d p_n}{dt} = \sum_j T_{mn} p_m, \quad T_{nm}(n\ne m) = t_{nm}, \quad T_{nn} = - \sum_{m \ne n} t_{nm}.
\end{equation}
The relaxation time is governed by eigenvalues of matrix $T_{nm}$. This matrix has one eigenvalue equal to zero (that correspond to the Bolzman distribution of probabilities). Its other eigenvalues are negative and correspond to some relaxation rates. To calculate the relaxation rate $\nu_i$ of transition between metastable states of the system that corresponds to the avalanche $i$ we expand the ``direction" of the relaxation (i.e., the vector of size $N_{st}$ that has only two nonzero elements $-1/\sqrt{2}$ and $1/\sqrt{2}$ corresponding to the initial state before the avalanche and the final state after the avalanche) over eigenvectors of $T_{nm}$. In this way we select only the eigenvectors of $T_{nm}$ that actually contribute to transitions between metastable states. From this vectors we select the one that corresponds to the minimal relaxation rate $\nu_{min}$. The relaxation time of the avalanche is then estimated as $\tau_i=1/\nu_{min}$ and the relaxation rate is $\nu_i = \nu_{min}$.

The found rate $\nu_i$ should be included into distribution $W(\ln(\nu))$ only if the discussed system can really be considered as independent (at least approximately) from other parts of Coulomb Glass. To prove it strictly one should directly calculate relaxation including all the sites of the sample. It is not possible in realistic time. Therefore, we check the two following issues before ascribing $\nu_i$ to the whole system (and including it into $W(\ln(\nu))$). Firstly we check that there are no possible 1e hops between the sites of the avalanche and external sites that go faster than $\tau_i$. Secondly we check that inclusion of any one external site into our computation do not alter the relaxation rate significantly (i.e. more then $25\%$). If both criteria are met we consider the avalanche as independent at given temperature (note that all discussed times depend on temperature) and include the rate $\nu_{i}$ to the distribution $W(\ln(\nu))$.

\begin{figure}[htbp]
    \centering
        \includegraphics[width=0.9\textwidth]{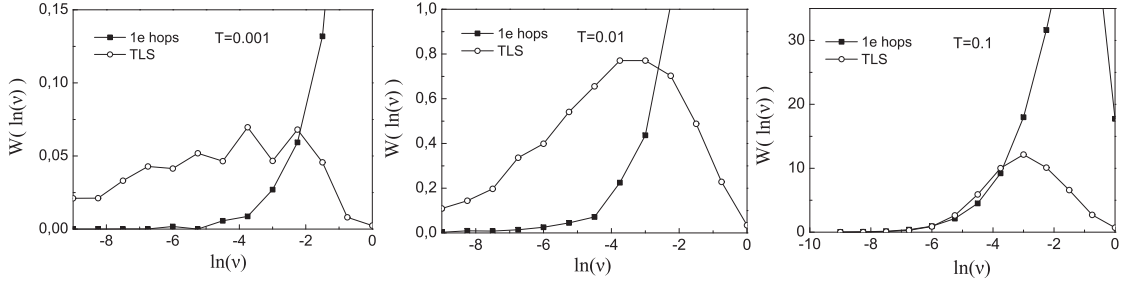}
        \caption{The weighted distribution $W(\ln(\nu))$ of relaxation rates of two-level systems (arb. units) compared to the  analogous distribution for 1e hops at different temperatures. The results are averaged over
        100 numerical samples.   }
    \label{fig:TLS1e}
\end{figure}

Figure \ref{fig:TLS1e} shows the weighted (with weights $w_T$) distribution $W(\ln(\nu))$ of logarithms of relaxation rates of two-level systems at different temperatures. To obtain this distribution we calculated all avalanches induced by dipole excitations in $100$ numerical samples, i.e., $100$ random site positions (the initial relaxation process for the samples was discussed above). This distribution is compared with the analogous distribution of rates of 1e hops. To get it we treated one-electron hops in the same way as two-level systems. We calculated the rate of every 1e hop (let us say that it is from site $i$ to the site $j$). Then we checked if the time of electron transition from site $i$ to site $j$ can be significantly changed by inclusion of any intermediate site. If it is not, then we included corresponding rate to our distribution with the weight  $w_T(\Delta E_{hop})$. $\Delta E_{hop}$ is the difference of energies before and after the hop.

It can be seen that 1e hops always dominate for high relaxation rates. However, at low temperatures there is a rate interval (with at least two orders of magnitude) where two-level systems dominate over one-electron hops. At higher temperatures ($T>0.1$) the part of the distribution of two level systems that corresponds to small relaxation rates dies out. Note that the temperature unit here is the Coulomb interaction of two electrons at the distance $1/\sqrt{n}$, where $n$ is 2D concentration. Thus $T=0.1$ is still much smaller then characteristic Coulomb interaction between neighboring sites. It is comparable with the characteristic energy $\Delta E$ of an avalanche (see Fig. \ref{fig:AvSE}).

It is interesting to compare our avalanches with metastable aggregates considered in \cite{agr1f,agrrel, agrOur0, agrour}. The metastable aggregates in discussion are chessboard clusters (may be with some extent of disorder). The electrons located on such clusters have two pseudoground states. The energies of these states are different only due to disorder and in any case are considered to be much lower than the energies of all other ``excited" states of the aggregate. In the absence of thermal excitation a transition between metastable states can occur only via co-tunneling of all electrons in the aggregate.

Let us assume that one of our avalanches corresponds to such a cluster and consider the energies of all its states (corresponding to different distribution of electrons over the sites participating in the avalanche). Then the two states with the lowest energies are the initial and the final states of the avalanche. It is not always so in our case. For example, some levels that can be considered as ``excited from the final state of the avalanche" can be lower than the initial state of the avalanche. Or vice versa: the states excited from the initial state can appear lower than the final state of the avalanche. It means that the relaxation of corresponding two-level system does not necessarily include co-tunneling of all the electrons of the avalanche even at $T=0$.

\begin{figure}[htbp]
    \centering
        \includegraphics[width=0.8\textwidth]{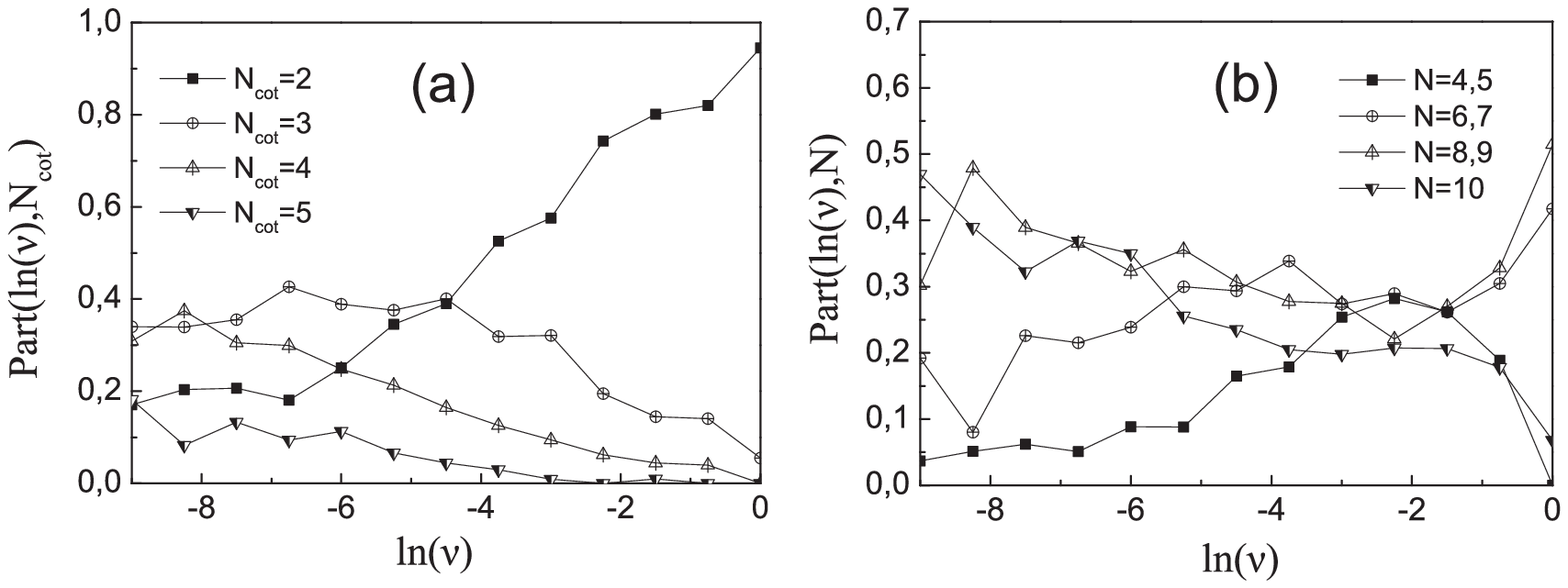}
        \caption{The participations of different $N_{cot}$ (a) and sizes $N$ of the avalanches (b) to the distribution $W(\ln(\nu))$. Temperature is $T=0.01$ }
    \label{fig:parts}
\end{figure}

Let us consider an avalanche with 10 sites and 5 electrons. Our algorithm ensures that all 1e transitions from both the initial and the final states of the avalanche increase the energy of the system and cannot take place at $T=0$. However, we cannot say the same about transitions of 2 or more electrons. It may occur (and often occurs) that to get from the initial state of the avalanche to its final state without thermal excitation one can move (for example) two electrons, then another one, and then the last two electrons. In this case the relaxation of this system at $T=0$ includes co-tunneling of only two electrons rather than all five electrons of the system. In this way we can define number $N_{cot}$ for every avalanche as a minimal number of electrons that should co-tunnel for the system to relax at $T=0$.

It is interesting which values of $N_{cot}$ determine the TLS-generated noise
at different frequencies. For this reason we calculated the participation ratios characterizing relative contributions
 of the metastable systems with different $N_{cot}$ to  the distribution $W(\ln(\nu))$:
\begin{equation}
{\rm Part}(\ln(\nu),N_{cot}) = \frac{W(\ln(\nu),N_{cot})}{W(\ln(\nu))}, \quad \sum_{N_{cot}}{\rm Part}(\ln(\nu),N_{cot}) = 1.
\end{equation}
Here $W(\ln(\nu))$ is the distribution of relaxation rates, $W(\ln(\nu),N_{cot})$ is the contribution to it from systems with given $N_{cot}$. In a similar way one can define participation ratios for different sizes $N$ of the avalanches, ${\rm Part}(\ln(f),N)$.
Figure \ref{fig:parts} shows these participation ratios  for temperature $T=0.01$.  It can be seen that $N_{cot} = 2$ dominates the distribution at high rates. However at lower rates larger $N_{cot}$ become important. The dependence is not that clear when size of an avalanche rather than $N_{cot}$ is considered.

Let us note that in our computations we were able to find relaxation rates only for avalanches with relatively small size $N \le 10$. The number of co-tunneling electrons, $N_{cot}$, is usually smaller than $N/2$. Therefore, we think that the contribution of large $N_{cot}$ is significantly underestimated in our computation.

 There is also the second reason to consider the obtained TLS contribution to the noise spectrum as underestimated. It can be shown that a two-level system with size $N=4$ should always be detected during an avalanche. However it is not the case for larger TLS. Let us consider a chessboard cluster from \cite{agr1f,agrrel,agrOur0, agrour} with size $N>4$. If one moves an electron from one site of the cluster to a neighboring one (as we do at the first step of the avalanche calculation) the cluster should relax to its second metastable state. However, it can occur that this relaxation cannot be achieved only with 1e hops with decrease of the total energy. In this case the discussed cluster will not be detected with a given avalanche. If the relaxation with 1e hops is impossible after any initial single-electron transition the cluster will not be detected with the avalanche method at all.

 In general our computation should be considered as a lower estimate for the noise spectrum of the Coulomb Glass. However, even this estimate shows that two-level systems dominate over 1e hops in the formation of noise spectrum at low frequencies and temperature.

\subsection{Effect of disorder}

Up to this moment we considered only idealistic situation when all the disorder in the system comes from random positions of the sites, i.e., there is no external random potential. However, in realistic systems a random potential is always present although one can try to make it small. In \cite{Efr-Ts} the evidence is present that the random potential comparable with the interaction between neighboring electrons significantly suppresses slow relaxation of conductivity. One may expect something similar in our calculations.

\begin{figure}[htbp]
    \centering
        \includegraphics[width=0.9\textwidth]{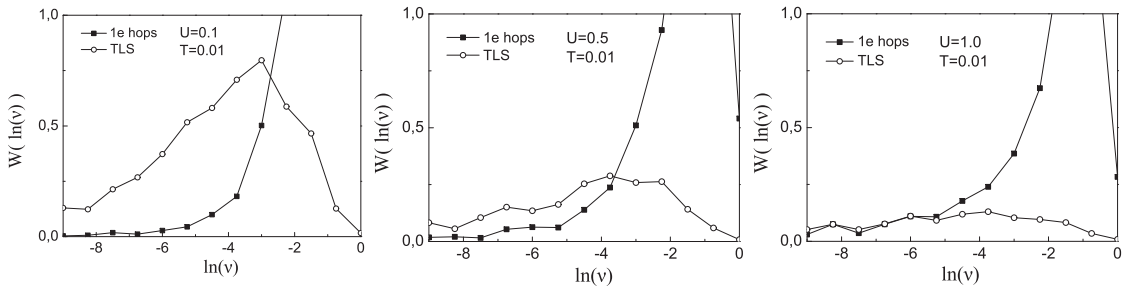}
        \caption{The distribution $W(\ln(\nu))$ for 1e hops and two-level systems at different disorder strengths for temperature $T=0.01$. The data is averaged over 50 disorder realizations. }
    \label{fig:disorder}
\end{figure}

In what follows we try to answer the question: how strong the random potential should be to suppress Coulomb Glass  phenomena? We consider numerical samples with various degree of random potential, i.e, with constant $U$ equal to $0.1$, $0.5$ and $1$. For each disorder strength we consider 50 numerical samples, i.e., 50 realizations of random site positions and random on-site energies. Figure \ref{fig:disorder} shows the results of computation (weighted distributions $W(\ln(\nu))$ of TLS and 1e hop relaxation rates) for considered values of disorder strength and temperature $T=0.01$. One can see that at $U=0.1$ the distributions are essentially the same as in the case of no energy disorder (Fig. \ref{fig:TLS1e}). At $U=0.5$ the part of TLS distribution, corresponding to low rates, is significantly suppressed. However, there remains a rate interval where it prevails over the distribution of 1e hop rates. Finally for $U=1$ one electron hops prevail at all the computed range of relaxation rates.

Thus one can see that to suppress low frequency noise generated by two-level systems the random potential should be comparable with the Coulomb interaction between neighboring sites.

\section{Conclusions}

We present a method based on the avalanche computation that allows to find metastable two-level systems in  numerical samples of the Coulomb Glass. We consider these systems as independent from the rest of the Coulomb Glass if (at the temperature under consideration) inclusion of any single external site does not significantly alter  the relaxation time of the system.

We computed the distribution of logarithms of relaxation rates of metastable systems that determined the contribution of these systems to the low-frequency noise. We compared this contribution with the contribution of 1e hops. It is shown that at low temperature and energy disorder, contribution of two-level systems to the noise spectrum is larger then contribution of one electron hops in a wide frequency range. Temperature of the order of characteristic energy of TLS and the random potential of the order of Coulomb interaction between neighboring sites suppress the TLS contribution to the low frequency noise.

Author is grateful to M. Goethe, Y.M. Galperin and J. Bergli for manu fruitful discussions. The work is supported by RFBR foundation
(grant for young scientists n. 12-02-31655)

\end{document}